\documentclass[reprint,amsmath,amssymb,aps]{revtex4-2}

\usepackage{graphicx}% Include figure files
\usepackage{dcolumn}% Align table columns on decimal point
\usepackage{bm}% bold math
\usepackage{hyperref}% add hypertext capabilities
\usepackage{xcolor}

\begin{document}
\newcommand{\eric}[1]{\textcolor{green}{[Eric: #1]}}
\newcommand{\hongzhe}[1]{\textcolor{magenta}{[HZ: #1]}}

\newcommand{\beq}{\begin{equation}}
\newcommand{\eeq}{\end{equation}}

\newcommand{\benum}{\begin{enumerate}}
\newcommand{\eenum}{\end{enumerate}}

\def\apj{ApJ}
\def\apjs{ApJS}
\def\aap{A \& Ap}
\def\pasp{PASP}
\def\pasj{PASJ}
\def\ssr{Sp Sci. Rev.}
\def\solphys{Solar Physics}
\def\mA{\mathcal A}
\def\araa{ARAA}
\def\nat{Nature}
\def\mnras{MNRAS}
\def\apj{ApJ}
\def\apjl{ApJL}
\def\aap{A\&A}
\def\apss{Ap\& Spac. Sup.}
\def\jcap{Journal of Cosmology and Astroparticle Physics}
\def\physrep{Physics Reports}
\def\pre{Physical Review E}
\def\physscr{Physica Scripta}
\def\aj{AJ}
\def\prl{Phys. Rev. Lett.}
\def\pnas{Proceedings of the National Academy of Science}
\def\qjras{Quarterly Journal of the Royal Astronomical Society}

\newcommand{\titlealpha}{\texorpdfstring{$\bm\alpha$}{p}}

\def\bar{\overline}
\newcommand{\abra}[1]{\left\langle{#1}\right\rangle}
\newcommand\remark[1]{{\color{red}[{\it{#1}}]}}
\newcommand{\edit}[2]{{{\color{cyan}\sout{#1}} {\color{brown}#2}}}

\newcommand{\bmA}{{\bm A}}
\newcommand{\bma}{{\bm a}}
\newcommand{\bmB}{{\bm B}}
\newcommand{\bmb}{{\bm b}}
\newcommand{\bmU}{{\bm U}}
\def\bmu{\bm{u}}
\newcommand{\bmJ}{{\bm J}}
\newcommand{\bmj}{{\bm j}}
\newcommand{\bmk}{{\bm{k}}}
\newcommand{\bmK}{{\bm{K}}}
\newcommand{\bmx}{{\bm{x}}}
\newcommand{\bmE}{{\bm{E}}}
\newcommand{\bme}{{\bm{e}}}

\newcommand{\del}{{\bm\nabla}}
\newcommand{\urms}{u_\text{rms}}
\newcommand{\kf}{k_\text{f}}
\newcommand{\kdis}{k_\text{dis}}
\newcommand{\bhel}{b_\text{hel}}
\newcommand{\uhel}{u_\text{hel}}
\newcommand{\alphaK}{\alpha_\text{k}}
\newcommand{\alphaM}{\alpha_\text{m}}
\newcommand{\ReN}{\text{Re}}
\newcommand{\RmN}{\text{Rm}}
\newcommand{\Rm}{\RmN}
\newcommand{\Pm}{\text{Pm}}
\newcommand{\knu}{k_\nu}
\newcommand{\Hk}{\mathcal{H}^\text{K}}
\newcommand{\Hc}{\mathcal{H}^\text{C}}
\newcommand{\Hm}{H^\text{M}}
\newcommand{\Hmi}{\Hm_i}
\newcommand{\Em}{E_\text{m}}
\newcommand{\tEm}{\tilde E_\text{m}}
\newcommand{\Ek}{E_\text{k}}
\newcommand{\kdiv}{k_\text{div}}
\newcommand{\chii}{\chi_\text{i}}
\newcommand{\chif}{\chi_\text{f}}
\newcommand{\tEhel}{\tilde{E}_\text{L}}
\newcommand{\Ma}{\text{Ma}}
\newcommand{\emf}{\mathcal{E}}
\newcommand{\cHm}{\mathcal{H}^\text{M}}
\newcommand{\alphazero}{\alpha_{0}}
\newcommand{\betazero}{\beta_{0}}
\newcommand{\kp}{k_\text{peak}}
\newcommand{\TD}{T_\text{D}}
\newcommand{\TR}{T_\text{R}}
\newcommand{\TDR}{{\TD,\TR}}
\newcommand{\dtt}[1]{\frac{\partial{#1}}{\partial\tilde t}}
\newcommand{\fssd}{f_\text{SSD}}

\preprint{}

\title{Helical dynamo growth at modest versus extreme magnetic Reynolds numbers}

\author{Hongzhe Zhou}
\email{hongzhe.zhou@sjtu.edu.cn}
\affiliation{%
Tsung-Dao Lee Institute, Shanghai Jiao Tong University,
800 Dongchuan Road, Shanghai 200240,
People's Republic of China}
\affiliation{%
Nordita, KTH Royal Institute of Technology and Stockholm University, Hannes Alfv\'ens v\"ag 12, SE-10691 Stockholm, Sweden}

\author{Eric G. Blackman}
\email{blackman@pas.rochester.edu}
\affiliation{%
Department of Physics and Astronomy, University of Rochester, Rochester, NY, 14627, USA}
\affiliation{%
Laboratory for Laser Energetics, University of Rochester, Rochester NY, 14623, USA}

\date{\today}

\begin{abstract}
Understanding large-scale magnetic field growth in astrophysical objects is a persistent challenge.
We tackle the long-standing question of how much helical large-scale dynamo growth occurs independent of the magnetic Reynolds number $(\Rm)$ in a closed volume.
From modest-$\Rm$ numerical simulations, we identify a pre-saturation regime when the large-scale field grows independently of $\Rm$, but to an $\Rm$-dependent magnitude.
For plausible magnetic spectra however, the analysis predicts the magnitude to be $\Rm$-independent and substantial as $\Rm\to\infty$.
This gives renewed optimism for the relevance of closed dynamos and pinpoints how modest $\Rm$ and hyper-diffusive simulations can cause misapprehension of the $\Rm\to\infty$ behavior.
\end{abstract}

\maketitle

\emph{Introduction.}---%
Large-scale magnetic fields of
many stars, planets, and galaxies 
require an \emph{in situ} dynamo mechanism to sustain against macroscopic and microscopic diffusion.
Plausible dynamo models involve long-lived fields produced by collective motions of stochastic turbulent eddies \cite{Parker1955},
often studied in the framework of mean-field electrodynamics \cite{Steenbeck+1966}.
Solutions to the suitably averaged mean-field induction equation
\beq
\partial_t\abra{\bmB}=\del\times\left(\abra{\bm U}\times\abra\bmB+\bm\emf\right)+\eta\nabla^2\abra{\bmB}
\label{eqn:induction}
\eeq
are sought, 
where $\bmB=\abra{\bmB}+\bmb$ is the total magnetic field measured in Alfv\'en units, $\abra{\ \cdot\ }$ is an average over a scale 
 assumed to be 
 much larger than the turbulent forcing scale,
and $\eta$ is the magnetic diffusivity.
We use lower case $\bm{b}$ to indicate the contribution to $\bmB$ with zero mean, and similar constructions for the magnetic vector potential $\bmA$ and velocity $\bmU$.
For statistically homogeneous and isotropic, kinetically helical turbulence,
the turbulent electromotive force (EMF) $\bm\emf$ contains a term $\alpha\abra{\bm B}$ that can amplify $\abra{\bm B}$ \citep{Steenbeck+1966,Moffatt1978,Raedler+2003,RaedlerStepanov2006}.
The increasing Lorentz force on the flow eventually quenches the dynamo.

A long-debated question is whether
the quenching becomes more severe at higher magnetic Reynolds numbers $\Rm$
\citep{KulsrudAnderson1992,CattaneoVainshtein1991,VainshteinCattaneo1992,
GruzinovDiamond1994,GruzinovDiamond1995,GruzinovDiamond1996,Ossendrijver+2001,TobiasCattaneo2013}.
In the ``dynamical quenching" (DQ) formalism, 
 gradients of $\abra{\bmB}$ are required for it to grow, and the dynamo quenching is controlled by the conservation
of magnetic helicity
\citep{Pouquet+1976,Kleeorin+1982,BlackmanField2000,BlackmanField2002,Blackman2003, Brandenburg+2008}.
In this formalism, significant growth of $\abra{\bmB}$
 occurs during an $\Rm$-independent regime, after which $\Rm$-dependent saturation occurs.
The field strength can reach super-equipartition values, but only on a resistively long time scale
\citep{BlackmanField2002,BrandenburgSarson2002,CandelaresiBrandenburg2013}.
Although resistive time scales are appropriate for planets, large $\Rm$ renders the
resistive growth time scale too long for many stellar and galactic contexts.
This raises the question of how strong $\abra{\bmB}$ gets before $\Rm$ dominates the evolution.

A substantial $\Rm$-independent regime has not yet been definitively identified in simulations. This, and the often impractically long time scale for the fully saturated phase, make it challenging to understand the origin of large-scale fields in
 astrophysical flows from helical dynamo.
Solutions to this problem include helicity fluxes \citep[e.g.,][]{BlackmanField2000,VishniacCho2001,SubramanianBrandenburg2004,HubbardBrandenburg2012,Rincon2021,GopalakrishnanSubramanian2022} and 
anisotropic forcing \citep{Bhat2022}.
See Ref. \cite{BrandenburgSubramanian2005} for a comprehensive review,
and also Refs. \cite{Brandenburg2018,Hughes2018,Rincon2019,BrandenburgNtormousi2022}. 

In this Letter, we investigate whether an $\Rm$-independent regime can exist before the very long resistive phase.
The helical dynamo in a closed system consists of three distinct temporal stages:
(i) a fast small-scale dynamo (SSD) that grows on turbulent time scales,
(ii) a large-scale dynamo (LSD) that is $\mathcal{O}(10)$ times slower than the SSD, and
(iii) a growth driven by magnetic helicity dissipation that operates on resistive time scales.
At very early times when the magnetic energy is negligible to the kinetic energy at all scales,
the SSD and LSD phases can be described in a unified framework \citep{Subramanian1999,Boldyrev+2005,Bhat+2016}.
Once the SSD nearly saturates, LSD takes over and potentially operates independently of resistivity
\citep{Graham+2012,Bhat+2019}.
The resistive phase dominates once the LSD saturates \citep{Brandenburg2001,BlackmanField2002,BrandenburgSarson2002,Bermudez+2022} 
Here we answer the question: does the LSD phase amplify $\abra{\bmB}$ to an $\Rm$-independent value before dynamical quenching transitions to the resistively limited asymptotic phase?
We call the regime whose $\Rm$
dependence we assess as the ``pre-quenched'' (PQ) regime, a definition to be further clarified later.

To answer our question from numerical simulations is challenging without careful interpretation, because the time scales of the LSD and the resistive phases are poorly separated at moderate values of $\Rm$ available. Instead of using time to delineate dynamo phases, here
we employ a new dynamo tracker which records how much the dynamo driver has been quenched.
At each quenching level, we analyze individually the $\Rm$-dependence of the LSD growth rate and the field strength. 
We then discuss the distinct implications of our analysis for the modest $\Rm$ values obtainable in the simulations versus the implications for asymptotically large $\Rm$.

\emph{Methods.}---%
We perform compressible magnetohydrodynamics simulations with an isothermal equation of state using the \textsc{Pencil Code} \citep{JOSS2021,zenodo}.
The velocity is driven in a $(2\pi)^3$-periodic box
using positively helical plane waves at a fixed forcing wave number $\kf$, 
but with random phases and directions at each time step.
The vector potential $\bm A$ is solved in the resistive gauge (i.e., the scalar potential is $\phi=-\eta\del\cdot\bm A$), but periodic boundary
conditions ensure that the magnetic helicity is gauge invariant.
For all runs, we use $\kf=4$ and Mach numbers $\text{Ma}\simeq0.1$.
The Reynolds numbers $\ReN=\urms/\nu\kf$ (with $\urms$ being the instantaneous root-mean-square velocity and $\nu$ being the viscosity) are kept roughly constant, $\simeq5$,
and the magnetic Prandtl number $\Pm=\nu/\eta$ is varied from $1$ to $100$.
This isolates the $\Rm$ dependence from that of $\ReN$.

We consider only the helical part of the magnetic field, as it is most relevant to the LSD.
Although the current helicity spectrum $\Hc$ is always gauge-independent,
 we formulate the equations using the magnetic helicity spectrum
$\cHm$ for convenience. For the present boundary conditions, the two are related by $\Hc=k^2\cHm$ where $k$ is the wave number.
We normalize energy and helicity spectra such that
 integration over all wave numbers yields the energy or helicity density.
We then decompose the large- and small-scale magnetic helicity densities as
\beq
\int \cHm_i\ \text{d}k=s_ik_i^{-1}\int k\left|\cHm_i\right|\ \text{d}k,
\label{eqn:hel_decomp}
\eeq
where $i=1,2$ denote large-scale ($k<\kf$) and small-scale ($k\geq\kf$)
modes respectively,
$s_i={\int \cHm_i\ \text{d}k}/{\int \left|\cHm_i\right|\ \text{d}k}$
is the mean handedness,
and
$k_i={\int k\left|\cHm_i\right|\ \text{d}k}/{\int \left|\cHm_i\right|\ \text{d}k}$
is the mean wave number of the helical fields.
Note that $s_i\in[-1,1]$ and $k_i>0$.
The non-dimensional energy density of the large-scale helical field is
$\tEhel=\urms^{-2}\int k|\cHm_1|\ \text{d}k$.
We define dimensionless time as
$\tilde t(t)=\int_0^t \urms(t')\kf\ \text{d}t'$
which is monotonic in $t$, and reduces to $\tilde t=t\urms\kf$ for constant $\urms$. 
We also define the dimensionless exponential growth rate $\tilde\gamma=\text{d}{\ln\tEhel}/\text{d}\tilde t$.

\emph{LSD growth rate versus $\Rm$}---%
For statistically isotropic and homogeneous turbulence, the turbulent EMF in Eq.~(\ref{eqn:induction}) takes the form
$\bm\emf=\alpha\abra\bmB-\beta\del\times\abra\bmB$.
The turbulent diffusivity is $\beta=\tau\urms^2/3$,
and $\tau=1/\urms\kf$.
In the DQ formalism, 
$\alpha=\alphaK+\alphaM$.
Here 
\beq
\alphaK=-\frac{1}{3}\abra{
\int_0^t\bmu(t)\cdot\del\times\bmu(t')\ \text{d}t'
}
\simeq-\frac{1}{3}\epsilon\urms
\label{eqn:alphak}
\eeq
is the the kinetic contribution, and the magnetic contribution
\beq
\alphaM=\frac{1}{3}\abra{
\int_0^t\bmb(t)\cdot\del\times\bmb(t')\ \text{d}t'
}\simeq
\frac{1}{3}\left(\int k\cHm_2\ \text{d}k\right)^{1/2}
\label{eqn:alpham}
\eeq
grows by DQ \cite{Pouquet+1976,Kleeorin+1982,BlackmanField2002} and is related to the small-scale current helicity multiplied by a correlation time.
In Eq.~(\ref{eqn:alphak}), we have assumed the velocity field to be fully helical, which is consistent with our simulations.
We also introduced a factor $\epsilon$ to capture the possible deviation of the correlation time between $\bmu$ and its curl, from the eddy turnover time $1/\urms\kf$.
Measuring $\epsilon$ can be method-dependent and so we treat it as a free parameter.
We shall see that $\epsilon=0.85$ is sufficient to explain simulations,
and more crucially, it has no influence on the implications on the $\Rm$-dependence of helical LSDs.

In the DQ formalism, $\alphaM$ grows in time, offsets $\alphaK$, and eventually quenches the dynamo.
We thus define $\chi\equiv-\epsilon\alphaM/\alphaK$ as the DQ factor, which is roughly the normalized current helicity. It contains no free parameter and is calculable from simulations.
The measured value of $\chi$ grows nearly monotonically in time with fluctuations.
In what follows, any quantity taken at $\chi=\chi'$ is meant to be its average
over the interval $\chi\in[\chi'-\delta',\chi'+\delta]$ with $\delta=\text{min}\left\{0.2\chi',0.05\right\}$,
unless otherwise specified.

For sufficiently large $\Rm$, SSD is excited at early times and the growth of the large-scale modes is dominated by nonlinear inter-mode interactions rather than the interaction with the velocity field through the $\alpha$ effect.
The value of $\chi$ at which the large-scale modes are dominated by LSD
is empirically found to be $\chi \simeq 0.1$,
as determined by two measurements:
(i) The mean-to-rms ratio, $\abra{B}^2/B_\text{rms}^2$,
remains constant at $\chi\leq0.1$ for all runs,
which is a signature SSD feature;
(ii) The SSD phase efficiently amplifies small-scale fields but with low fractional magnetic helicity, e.g. an average value of $0.05$ for the simulation run at $\Rm=259$.
Thus $\chi >0.1$ is primarily a result of the helical LSD.

\begin{figure*}
\centering
\includegraphics[width=1.8\columnwidth]{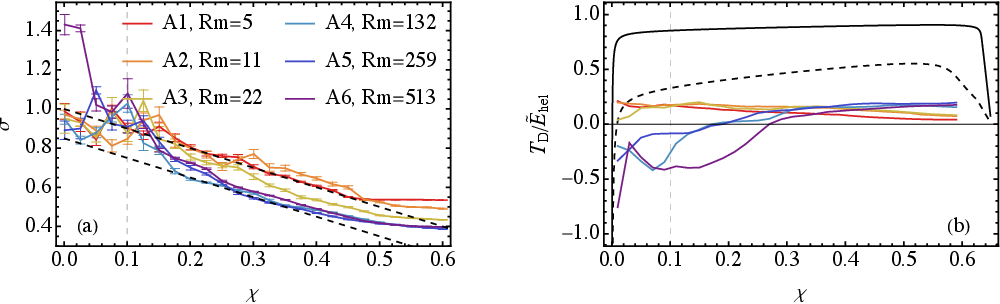}
\caption{$\Rm$-independent growth and $\Rm$-dependent field strength in helical LSDs.
In both panels, the vertical dashed line at $\chi=0.1$ indicates the start of the LSD phase.
(a) Measured values of LSD efficiency $\sigma$ [Eq.~(\ref{eqn:sigma})]
versus the dynamical quenching factor $\chi$ for different $\Rm$.
The two black dashed lines indicate the theoretical expectation $\sigma^\text{th}=\epsilon-\chi$
with different ratios of the time scales of $\alphaK$ and $\beta$: $\epsilon=1$ (upper) and $\epsilon=0.85$ (lower).
The LSD growth rate becomes independent of $\Rm$ for runs A4 to A6, i.e., when $\Rm\geq 130$.
(b) Fractional contributions to $\tEhel$ from the dynamical term, ${\TD}/\tEhel$,
from the simulation data (colored) and the theoretical LSD model (black).
Hence ${\TR}/\tEhel=1-\TD/\tEhel$ decreases with increasing $\Rm$ but still dominates when $\Rm\simeq500$,
leading to $\Rm$-dependent field strengths in simulations.
Two theoretical LSD curves are shown for $\Rm=2500$ with $q=8/3>2$ (solid) and $q=1<2$ (dashed).}
\label{fig:sigma_chi_TDR}
\end{figure*}

For a helical LSD without a mean flow (i.e., an $\alpha^2$ dynamo),
the energy growth rate of the mode at wave number $k_1$ is
$\gamma=2|\alpha|k_1-2(\beta+\eta)k_1^2$ \citep{BrandenburgSubramanian2005}.
Using the eddy turnover rate $\urms\kf$ for normalization, we have \citep{sm}
\beq
\tilde\gamma=\frac{\gamma}{\urms\kf}=
\frac{2k_1}{3\kf}(\epsilon-\chi)-\left(\frac{2}{3}+\frac{2}{\Rm}\right)\left(\frac{k_1}{\kf}\right)^2,
\label{eqn:gamma_chi}
\eeq
where $\Rm=\urms/\eta\kf$ is the instantaneous magnetic Reynolds number.
The LSD initially operates kinematically when $\chi\ll1$,
but is then dynamically quenched by $\chi$ due to the growing small-scale
current helicity.
The maximal value $\chi$ can obtain is analytically determined by $\tilde\gamma=0$ to be $\epsilon-k_1/\kf$ when $\Rm\to\infty$.
For the runs with large $\Rm$ (A4 to A6), we have approximately $\epsilon=0.85$ (see Fig.~\ref{fig:sigma_chi_TDR}(a) and also the next paragraph),
and hence the LSD regime for which $\chi\in[0.1,0.6)$ 
quantitatively demarks the PQ regime whose $\Rm$ dependence we will assess.
Resistive diffusion of magnetic helicity reduces the growth
rate of $\chi$ and thus slows the magnetic back-reaction on the LSD, but does not directly show up in the LSD growth rate.
Hence at any given $\chi$ during the LSD phase, the LSD growth rate can become $\Rm$-independent at lower $\Rm$ than the value at which the mean-field strength becomes $\Rm$-independent.
We focus on $\tilde\gamma$ here and discuss the influence of resistivity on the field strengths later.

Eq.~(\ref{eqn:gamma_chi}) is the theoretical expectation of the LSD growth rate.
To validate it from simulations, it is more convenient to define an LSD efficiency $\sigma=\epsilon-\chi$ and express this in terms of $\tilde\gamma$ using Eq.~(\ref{eqn:gamma_chi}),
\beq
\sigma=\frac{3\kf}{2k_1}\tilde\gamma+\left(1+\frac{3}{\Rm}\right)\frac{k_1}{\kf}.
\label{eqn:sigma}
\eeq
The right side is directly measured in simulations as plotted in Fig.~\ref{fig:sigma_chi_TDR}(a),
with varying $\Rm$ for different colors.
Overall, the measured values of $\sigma$ at $\chi\in[0.1,0.5]$ can be well described by the theoretical expectation $\epsilon-\chi$,
with $\epsilon\in[0.85,1]$, as indicated by the two black dashed lines.
This implies a modest $15\%$ deficit in the correlation time between $\bmu$ and its curl (which enters $\alphaK$) than that between $\bmu$ and itself (which enters $\beta$),
and might explain the sub-maximal LSD efficiency of Ref. \citep{Bhat+2019}.

The agreement between measured values and theoretical expectation of $\sigma$
validates Eq.~(\ref{eqn:gamma_chi}),
and thus justifies that
(i) the DQ formalism correctly describes the $\alpha^2$ dynamo, and
(ii) the LSD growth rate is asymptotically independent of $\Rm$ when $\Rm\gg1$.
To see the latter point, notice that on the right of Eq.~(\ref{eqn:gamma_chi}) the only two $\Rm$-dependent quantities are
$\Rm$ itself and $k_1$.
The $k_1$ is initially the value $\sim \kf/2$ which maximizes $\tilde\gamma$.
Later, $k_1$ decreases to the lowest wave number available in the system, and this evolution may depend on $\Rm$.
But overall, $k_1$ always changes by a factor of $\kf L/4\pi$ if $L$ is the length scale of the system, which is $\Rm$-independent.
Hence the $\Rm$-dependence of $\tilde\gamma$ introduced by $k_1$ is quite weak and $\tilde\gamma$ will not decrease to some resistively small value.
Indeed, Fig.~\ref{fig:sigma_chi_TDR}(a) shows that $\sigma$, and hence $\tilde\gamma$, become $\Rm$-independent once $\Rm\gtrsim130$.
An $\Rm$-independent growth rate was previously reported at $\Rm\gtrsim500$ \citep{Bhat+2019},
but here we provide evidence at lower $\Rm$.

\emph{$\Rm$-dependent field strengths in simulations}---
Although the LSD growth rate is verified to be independent of $\Rm$ for $\Rm\rightarrow\infty$, 
the simulated values of $\tEhel$ decrease with increasing $\Rm$ at fixed $\chi$.
To quantify this, we define an exponent $p(\chi)$ by fitting the power-law $\tEhel(\chi,\Rm)\propto \Rm^{p(\chi)}$.
We find that $p\simeq-0.5$ for all $\chi\in[0.01,0.6]$,
indicating that during the SSD phase and at all stages of the LSD, the mean-field energy decreases with increasing $\Rm$.
We now explain this dependence.

$\tEhel$ can be inferred from the total magnetic helicity
without integrating the growth equation.
Consider the case where the volume-averaged magnetic helicity is zero initially, 
but later gains $\Delta H=\Hm_1 + \Hm_2$ due to resistive diffusion,
where $\Hm_{1,2}$ are the average magnetic helicity of the large- and small-scale fields, respectively.
Using Eq.~(\ref{eqn:hel_decomp}) we have
$\Hm_1=s_1k_1^{-1}\tEhel\urms^2$, and therefore
\beq
\tEhel(\chi,\Rm)=-\frac{k_1}{s_1}\frac{\Hm_2}{\urms^2}
+\frac{k_1\Delta H}{s_1\urms^2}.
\label{eqn:helicity_conservation}
\eeq
We denote the two terms on the right of Eq.~(\ref{eqn:helicity_conservation}) by ${\TD}$ and ${\TR}$, respectively,
so that $\tEhel={\TD}+{\TR}$.
Note that ${\TR}$ is from the resistive loss of magnetic helicity but ${\TD}$ is purely dynamic.
The resistive term does not amplify the large-scale field directly,
but slows the growth of $\chi$,
thereby weakening the back-reaction and allowing more large-scale growth.
The ratio $\TD/\tEhel$ determines whether resistive effects dominate and this is to be assessed below.

For all runs at all times, the magnetic fields near the resistive wave number
have positive magnetic helicity, whilst those at the lowest wave numbers have negative helicity.
Hence $\Delta H<0$ and $s_1<0$ always.
As we discuss in detail in Supplemental Material \citep{sm}, $\Hm_2<0$ during the SSD phase, and hence $\TD/\tEhel<0$ initially.
In what follows we focus on the LSD phase when $\Hm_2>0$, the period during which $\TD$ and $\TR$ are both positive.

The evolution of ${\TD}/\tEhel$ is shown in Fig.~\ref{fig:sigma_chi_TDR}(b).
The LSD starts to dominate the field growth at $k_1$ at $\chi \gtrsim 0.1$,
and its back-reaction on the small scales grows ${\TD}$.
By the time $\chi=0.6$ which is close to the end of the LSD regime, 
Eq.~(\ref{eqn:helicity_conservation}) determines how much the LSD has benefited from resistive contributions.
That ${\TD}<{\TR}$ 
implies that the LSD quenching is still weakened
substantially by the resistive dissipation of small-scale current helicity,
and therefore the
PQ regime depends strongly on $\Rm$.
This is why $\tEhel$ decreases with increasing $\Rm$ at fixed $\chi$.

\emph{Implications for higher $\Rm$.}---%
Fig. \ref{fig:sigma_chi_TDR}(b) shows that by $\chi=0.6$, the fractional contribution from ${\TD}$ increases with increasing $\Rm$, but stops increasing for the highest $\Rm$ run A6.
We now describe why this apparent saturation may not apply for much larger $\Rm$, and why ${\TD}$ might actually dominate at $\chi=0.6$ as $\Rm\rightarrow \infty$ and become $\Rm$-independent.

Since ${\TR}\propto\Delta H$ and is negligible during the LSD phase in the $\Rm\to\infty$ limit, the necessary condition for an $\Rm$-independent PQ
regime is $\text{d}{\TD}/\text{d} \Rm \rightarrow 0$ as $\Rm\rightarrow \infty$.
In the LSD phase, the small-scale magnetic helicity spectrum $\cHm_2$ is of one sign, 
so we write ${\TD}=-s_2k_1\chi^2/s_1k_2$.
Since $|s_{1,2}|\simeq1$ and $k_1$ is bounded from below, an
$\Rm$-independent PQ regime requires 
$k_2$ to depend at most weakly on $\Rm$ at fixed $\chi$,
which is determined by the magnetic helicity spectrum
as explained below.

Consider a magnetic helicity spectrum, $\cHm_2(k)\propto k^{-q}$ in the inertial range.
This is appropriate for $\Pm<1$ flows.
Using Eq.~(\ref{eqn:hel_decomp}) for $k_2$, we then have
that 
$k_2/\kf=F(q-1)/F(q)$,
%$\kf/k_2=F(q)/F(q-1)$
where $F(x)=\int_1^r x^{-q}\ \text{d}x$,
and $r$ is the ratio between the dissipative scale of the helical fields and $\kf$.
When $\Rm\to\infty$, we have $r\gg1$,
so that $k_2/k_f=(q-1)/(q-2)$ when $q>2$, but diverges for $q\leq 2$.
%\beq
%\frac{\kf}{k_2}\to\left\{
%\begin{aligned}
%&\frac{q-2}{q-1}, &q>2,\\
%&\frac{1}{\ln r}, &q=2,\\r
%&\frac{2-q}{q-1}\frac{1}{r^{2-q}}, &1<q<2,\\
%&\frac{\ln r}{r}, &q=1,\\
%&\frac{2-q}{1-q}\frac{1}{r}, &q<1
%\end{aligned}
%\right\}.
%\eeq
%
Hence an $\Rm$-independent PQ regime arises if $q>2$.

For $\Pm>1$ flows whose magnetic energy and helicity spectra may have broken power laws at $k\geq\kf$, the conditions for
an $\Rm$-independent PQ regime become
(i) a $q>2$ range exists, and 
(ii) the wave number above which $q>2$ does not increase with increasing $\Rm$.
Such evidence is indeed observed from our $\Pm\geq1$ simulations.
For the two highest-$\Rm$ runs (at $\Rm\simeq250$ and $500$), we find that the wave numbers at which $q-2$ changes sign are both $\simeq2\kf$ and do not scale with $\Rm$.
This is consistent with previous indications that the peak wave number of the magnetic energy spectrum for large-$\Pm$ SSDs remains $\Rm$-independent for large $\Rm$ from both theory
\cite{Subramanian1999} and simulation \cite{Galishnikova+2022}.
Hence, our simulations imply an $\Rm$-independent PQ regime for $\Pm\geq1$ flows.

Fig. \ref{fig:sigma_chi_TDR}(b) shows two theoretical predictions for the fractional dynamical contribution $\TD/\tEhel$ at $\Rm=2500$ in black, using a three-scale model (see Supplemental Material \citep{sm}),
with $q=8/3$ implying $\Rm$-independent quenching and $q=1$ implying $\Rm$-dependent quenching.
For the latter, the resistive terms still contributes nearly half of the large-scale helical field, implying an $\Rm$-dependent field strength during the PQ phase.
Near the fully quenched state at $\chi\simeq\epsilon-k_1/\kf$, the LSD growth rate has become smaller than the resistive loss rate of the small-scale magnetic helicity, and therefore the growth of the large-scale field is always dominated by the resistive term for any finite $\Rm$.
This leads to the decay of $\TD/\tEhel$ at large $\chi$ for both theoretical curves in Fig. \ref{fig:sigma_chi_TDR}(b).

The spectrum may also evolve from shallower than the aforementioned threshold at early times to steeper at later times. Then the influence of $\Rm$ on the saturated state could still be small, as is crudely suggested in a four-scale approach \citep{Blackman2003}.
Future high-resolution simulations for both $\Pm>1$ and $\Pm<1$ are needed to concretely confirm the spectral slope of the magnetic helicity at large $\Rm$ and its temporal evolution.

To summarize, for a magnetic helicity spectrum that satisfies the conditions mentioned three paragraphs above,
the $\Rm$-independent value that $\tEhel$ can obtain at any $\chi$ is
\beq
\lim_{\Rm\to\infty}\tEhel(\chi,\Rm)
=\lim_{\Rm\to\infty} {\TD}
=-\frac{s_2}{s_1}\frac{k_1}{\kp}\frac{q-2}{q-1}\chi^2,
\label{eqn:tEhel_sat}
\eeq
where $\kp$ is the $\Rm$-independent peak of the magnetic helicity spectrum
and $q>2$ is the slope at $k\geq\kp$.
Eq.~(\ref{eqn:tEhel_sat}) is the lower bound for any case with finite $\Rm$, to which the positive ${\TR}$ term will additionally contribute.
Since we have shown that $\tilde\gamma$ is asymptotically independent of $\Rm$,
it follows that the time to reach this lower bound is also $\Rm$-independent when $\Rm\to\infty$.
For all of our simulation runs, 
we observe $\tEhel$ is more than $12.5$ times this lower bound by taking $s_1/s_2=-1$, $\kp=2\kf$ and $q=8/3$, again highlighting the dominance of the resistive contribution. 

Furthermore, assuming $s_1/s_2=-1$, $\kp=2\kf$, $q=8/3$, $\kf/k_1=5$ and $\chi^2=1-k_1/\kf$, we find this lower bound to be $E_\text{L}/\urms^2\simeq \abra{B}^2/\abra{b^2}=0.032$, comparable to some observed galactic magnetic fields \citep{Beck+2019}
which have benefited from the $\Rm$-independent $\Omega$ effect and possible helicity fluxes.
Hence, the DQ formalism predicts a substantial lower bound for the large-scale magnetic energy.

\emph{Conclusions.}---%
Focusing on the PQ regime, namely the growth after the SSD saturates but before the resistive phase, our simulations and analyses reveal that:
(i) For isotropically helically forced flows, the large-scale field growth rate becomes $\Rm$-independent at modest $\Rm$ in accordance with the DQ formalism. 
Namely, $\alphaK$ remains $\Rm$-independent during the course of the simulation and 
the Lorentz-force back-reaction that ends the LSD regime exerts itself not by suppressing $\alphaK$, but by growth of $\alphaM$.
(ii) In contrast, the large-scale field strength attained in the PQ regime for $\Rm$ values accessible in the simulations is $\Rm$-dependent, being dominated by the resistive loss of magnetic helicity even for
$\Rm\simeq500$.
(iii) However, the same LSD analysis shows that the $\Rm\rightarrow \infty$ dependence
of the PQ regime depends on the evolution of the current helicity spectrum, or equivalently the magnetic helicity spectrum given the boundary conditions. For sufficiently steep magnetic helicity spectra, the regime becomes $\Rm$-independent as $\Rm\rightarrow \infty$, with the large-scale magnetic energy lower bound given by Eq.~(\ref{eqn:tEhel_sat}).

These results imply that when 
the current helicity spectrum falls off more steeply than $k^{0}$, efficient LSD growth is possible in high-$\Rm$ $\alpha^2$ or $\alpha^2$-$\Omega$ dynamos of stars and galaxies. This applies even without boundary helicity fluxes, although systems requiring fast cycle periods would favor 
helicity flux driven dynamos. 
For shallower spectra, helicity fluxes
or some non-helically driven LSD \citep[e.g.][]{Skoutnev+2022} would be need to explain the observed field strengths, let alone fast cycle periods. 
For planetary dynamos whose resistive time scales can be comparable to LSD dynamical
times, $\alpha$ quenching is significantly weakened by resistive diffusion and so the LSD is much less constrained by the slope of the current helicity spectrum.

Finally, hyper-diffusivity is sometimes used for mimicking high-$\Rm$ flows
\citep{BrandenburgSarson2002}.
Because the dissipation rate depends strongly on wave number,
magnetic energy piles up near the resistive scale (bottleneck effect)
\citep{Biskamp+1998}.
If these resistive-scale fields are helical,
$k_2$ becomes large and our analysis shows that the 
LSD growth rate 
is strongly quenched
even though it eventually leads to super-equipartition magnetic energies.
Hence, helical LSDs with hyper-diffusion are actually less effective for inferring realistic asymptotic LSD behavior.

Nordita is sponsored by Nordforsk.
We acknowledge allocation of computing resources from the
Swedish National Allocations Committee at the Center for Parallel
Computers at the Royal Institute of Technology in Stockholm and
Link\"oping. EB acknowledges the Isaac Newton Institute for Mathematical Sciences, Cambridge, for support and hospitality during the programme ``Frontiers in dynamo theory: from the Earth to the stars''. This work was supported by EPSRC grant no EP/R014604/1.
EB acknowledges support from grants US Department of Energy DE-SC0001063, DE-SC0020432, DE-SC0020103, and US NSF grants AST-1813298, PHY-2020249.

Data and post-processing programs for this article are available on Zenodo at doi:10.5281/zenodo.7632994 \citep{zenodo}.

\bibliography{refs}

\begin{thebibliography}{46}%
\makeatletter
\providecommand \@ifxundefined [1]{%
 \@ifx{#1\undefined}
}%
\providecommand \@ifnum [1]{%
 \ifnum #1\expandafter \@firstoftwo
 \else \expandafter \@secondoftwo
 \fi
}%
\providecommand \@ifx [1]{%
 \ifx #1\expandafter \@firstoftwo
 \else \expandafter \@secondoftwo
 \fi
}%
\providecommand \natexlab [1]{#1}%
\providecommand \enquote  [1]{``#1''}%
\providecommand \bibnamefont  [1]{#1}%
\providecommand \bibfnamefont [1]{#1}%
\providecommand \citenamefont [1]{#1}%
\providecommand \href@noop [0]{\@secondoftwo}%
\providecommand \href [0]{\begingroup \@sanitize@url \@href}%
\providecommand \@href[1]{\@@startlink{#1}\@@href}%
\providecommand \@@href[1]{\endgroup#1\@@endlink}%
\providecommand \@sanitize@url [0]{\catcode `\\12\catcode `\$12\catcode
  `\&12\catcode `\#12\catcode `\^12\catcode `\_12\catcode `\%12\relax}%
\providecommand \@@startlink[1]{}%
\providecommand \@@endlink[0]{}%
\providecommand \url  [0]{\begingroup\@sanitize@url \@url }%
\providecommand \@url [1]{\endgroup\@href {#1}{\urlprefix }}%
\providecommand \urlprefix  [0]{URL }%
\providecommand \Eprint [0]{\href }%
\providecommand \doibase [0]{https://doi.org/}%
\providecommand \selectlanguage [0]{\@gobble}%
\providecommand \bibinfo  [0]{\@secondoftwo}%
\providecommand \bibfield  [0]{\@secondoftwo}%
\providecommand \translation [1]{[#1]}%
\providecommand \BibitemOpen [0]{}%
\providecommand \bibitemStop [0]{}%
\providecommand \bibitemNoStop [0]{.\EOS\space}%
\providecommand \EOS [0]{\spacefactor3000\relax}%
\providecommand \BibitemShut  [1]{\csname bibitem#1\endcsname}%
\let\auto@bib@innerbib\@empty
%</preamble>
\bibitem [{\citenamefont {{Parker}}(1955)}]{Parker1955}%
  \BibitemOpen
  \bibfield  {author} {\bibinfo {author} {\bibfnamefont {E.~N.}\ \bibnamefont
  {{Parker}}},\ }\bibfield  {title} {\bibinfo {title} {{Hydromagnetic Dynamo
  Models.}},\ }\href {https://doi.org/10.1086/146087} {\bibfield  {journal}
  {\bibinfo  {journal} {\apj}\ }\textbf {\bibinfo {volume} {122}},\ \bibinfo
  {pages} {293} (\bibinfo {year} {1955})}\BibitemShut {NoStop}%
\bibitem [{\citenamefont {{Steenbeck}}\ \emph {et~al.}(1966)\citenamefont
  {{Steenbeck}}, \citenamefont {{Krause}},\ and\ \citenamefont
  {{R{\"a}dler}}}]{Steenbeck+1966}%
  \BibitemOpen
  \bibfield  {author} {\bibinfo {author} {\bibfnamefont {M.}~\bibnamefont
  {{Steenbeck}}}, \bibinfo {author} {\bibfnamefont {F.}~\bibnamefont
  {{Krause}}},\ and\ \bibinfo {author} {\bibfnamefont {K.~H.}\ \bibnamefont
  {{R{\"a}dler}}},\ }\bibfield  {title} {\bibinfo {title} {{Berechnung der
  mittleren LORENTZ-Feldst{\"a}rke f{\"u}r ein elektrisch leitendes Medium in
  turbulenter, durch CORIOLIS-Kr{\"a}fte beeinflu{\ss}ter Bewegung}},\ }\href
  {https://doi.org/10.1515/zna-1966-0401} {\bibfield  {journal} {\bibinfo
  {journal} {Zeitschrift Naturforschung Teil A}\ }\textbf {\bibinfo {volume}
  {21}},\ \bibinfo {pages} {369} (\bibinfo {year} {1966})}\BibitemShut
  {NoStop}%
\bibitem [{\citenamefont {{Moffatt}}(1978)}]{Moffatt1978}%
  \BibitemOpen
  \bibfield  {author} {\bibinfo {author} {\bibfnamefont {H.~K.}\ \bibnamefont
  {{Moffatt}}},\ }\href@noop {} {\emph {\bibinfo {title} {{Magnetic field
  generation in electrically conducting fluids}}}}\ (\bibinfo  {publisher}
  {Cambridge University Press},\ \bibinfo {year} {1978})\BibitemShut {NoStop}%
\bibitem [{\citenamefont {{R{\"a}dler}}\ \emph {et~al.}(2003)\citenamefont
  {{R{\"a}dler}}, \citenamefont {{Kleeorin}},\ and\ \citenamefont
  {{Rogachevskii}}}]{Raedler+2003}%
  \BibitemOpen
  \bibfield  {author} {\bibinfo {author} {\bibfnamefont {K.-H.}\ \bibnamefont
  {{R{\"a}dler}}}, \bibinfo {author} {\bibfnamefont {N.}~\bibnamefont
  {{Kleeorin}}},\ and\ \bibinfo {author} {\bibfnamefont {I.}~\bibnamefont
  {{Rogachevskii}}},\ }\bibfield  {title} {\bibinfo {title} {{The Mean
  Electromotive Force for MHD Turbulence: The Case of a Weak Mean Magnetic
  Field and Slow Rotation}},\ }\href
  {https://doi.org/10.1080/0309192031000151212} {\bibfield  {journal} {\bibinfo
   {journal} {Geophysical and Astrophysical Fluid Dynamics}\ }\textbf {\bibinfo
  {volume} {97}},\ \bibinfo {pages} {249} (\bibinfo {year} {2003})},\ \Eprint
  {https://arxiv.org/abs/astro-ph/0209287} {arXiv:astro-ph/0209287 [astro-ph]}
  \BibitemShut {NoStop}%
\bibitem [{\citenamefont {{R{\"a}dler}}\ and\ \citenamefont
  {{Stepanov}}(2006)}]{RaedlerStepanov2006}%
  \BibitemOpen
  \bibfield  {author} {\bibinfo {author} {\bibfnamefont {K.-H.}\ \bibnamefont
  {{R{\"a}dler}}}\ and\ \bibinfo {author} {\bibfnamefont {R.}~\bibnamefont
  {{Stepanov}}},\ }\bibfield  {title} {\bibinfo {title} {{Mean electromotive
  force due to turbulence of a conducting fluid in the presence of mean
  flow}},\ }\href {https://doi.org/10.1103/PhysRevE.73.056311} {\bibfield
  {journal} {\bibinfo  {journal} {\pre}\ }\textbf {\bibinfo {volume} {73}},\
  \bibinfo {eid} {056311} (\bibinfo {year} {2006})},\ \Eprint
  {https://arxiv.org/abs/physics/0512120} {arXiv:physics/0512120
  [physics.flu-dyn]} \BibitemShut {NoStop}%
\bibitem [{\citenamefont {{Kulsrud}}\ and\ \citenamefont
  {{Anderson}}(1992)}]{KulsrudAnderson1992}%
  \BibitemOpen
  \bibfield  {author} {\bibinfo {author} {\bibfnamefont {R.~M.}\ \bibnamefont
  {{Kulsrud}}}\ and\ \bibinfo {author} {\bibfnamefont {S.~W.}\ \bibnamefont
  {{Anderson}}},\ }\bibfield  {title} {\bibinfo {title} {{The Spectrum of
  Random Magnetic Fields in the Mean Field Dynamo Theory of the Galactic
  Magnetic Field}},\ }\href {https://doi.org/10.1086/171743} {\bibfield
  {journal} {\bibinfo  {journal} {\apj}\ }\textbf {\bibinfo {volume} {396}},\
  \bibinfo {pages} {606} (\bibinfo {year} {1992})}\BibitemShut {NoStop}%
\bibitem [{\citenamefont {{Cattaneo}}\ and\ \citenamefont
  {{Vainshtein}}(1991)}]{CattaneoVainshtein1991}%
  \BibitemOpen
  \bibfield  {author} {\bibinfo {author} {\bibfnamefont {F.}~\bibnamefont
  {{Cattaneo}}}\ and\ \bibinfo {author} {\bibfnamefont {S.~I.}\ \bibnamefont
  {{Vainshtein}}},\ }\bibfield  {title} {\bibinfo {title} {{Suppression of
  Turbulent Transport by a Weak Magnetic Field}},\ }\href
  {https://doi.org/10.1086/186093} {\bibfield  {journal} {\bibinfo  {journal}
  {\apjl}\ }\textbf {\bibinfo {volume} {376}},\ \bibinfo {pages} {L21}
  (\bibinfo {year} {1991})}\BibitemShut {NoStop}%
\bibitem [{\citenamefont {{Vainshtein}}\ and\ \citenamefont
  {{Cattaneo}}(1992)}]{VainshteinCattaneo1992}%
  \BibitemOpen
  \bibfield  {author} {\bibinfo {author} {\bibfnamefont {S.~I.}\ \bibnamefont
  {{Vainshtein}}}\ and\ \bibinfo {author} {\bibfnamefont {F.}~\bibnamefont
  {{Cattaneo}}},\ }\bibfield  {title} {\bibinfo {title} {{Nonlinear
  Restrictions on Dynamo Action}},\ }\href {https://doi.org/10.1086/171494}
  {\bibfield  {journal} {\bibinfo  {journal} {\apj}\ }\textbf {\bibinfo
  {volume} {393}},\ \bibinfo {pages} {165} (\bibinfo {year}
  {1992})}\BibitemShut {NoStop}%
\bibitem [{\citenamefont {{Gruzinov}}\ and\ \citenamefont
  {{Diamond}}(1994)}]{GruzinovDiamond1994}%
  \BibitemOpen
  \bibfield  {author} {\bibinfo {author} {\bibfnamefont {A.~V.}\ \bibnamefont
  {{Gruzinov}}}\ and\ \bibinfo {author} {\bibfnamefont {P.~H.}\ \bibnamefont
  {{Diamond}}},\ }\bibfield  {title} {\bibinfo {title} {{Self-consistent theory
  of mean-field electrodynamics}},\ }\href
  {https://doi.org/10.1103/PhysRevLett.72.1651} {\bibfield  {journal} {\bibinfo
   {journal} {\prl}\ }\textbf {\bibinfo {volume} {72}},\ \bibinfo {pages}
  {1651} (\bibinfo {year} {1994})}\BibitemShut {NoStop}%
\bibitem [{\citenamefont {{Gruzinov}}\ and\ \citenamefont
  {{Diamond}}(1995)}]{GruzinovDiamond1995}%
  \BibitemOpen
  \bibfield  {author} {\bibinfo {author} {\bibfnamefont {A.~V.}\ \bibnamefont
  {{Gruzinov}}}\ and\ \bibinfo {author} {\bibfnamefont {P.~H.}\ \bibnamefont
  {{Diamond}}},\ }\bibfield  {title} {\bibinfo {title} {{Self-consistent mean
  field electrodynamics of turbulent dynamos}},\ }\href
  {https://doi.org/10.1063/1.871495} {\bibfield  {journal} {\bibinfo  {journal}
  {Physics of Plasmas}\ }\textbf {\bibinfo {volume} {2}},\ \bibinfo {pages}
  {1941} (\bibinfo {year} {1995})}\BibitemShut {NoStop}%
\bibitem [{\citenamefont {{Gruzinov}}\ and\ \citenamefont
  {{Diamond}}(1996)}]{GruzinovDiamond1996}%
  \BibitemOpen
  \bibfield  {author} {\bibinfo {author} {\bibfnamefont {A.~V.}\ \bibnamefont
  {{Gruzinov}}}\ and\ \bibinfo {author} {\bibfnamefont {P.~H.}\ \bibnamefont
  {{Diamond}}},\ }\bibfield  {title} {\bibinfo {title} {{Nonlinear mean field
  electrodynamics of turbulent dynamos}},\ }\href
  {https://doi.org/10.1063/1.871981} {\bibfield  {journal} {\bibinfo  {journal}
  {Physics of Plasmas}\ }\textbf {\bibinfo {volume} {3}},\ \bibinfo {pages}
  {1853} (\bibinfo {year} {1996})}\BibitemShut {NoStop}%
\bibitem [{\citenamefont {{Ossendrijver}}\ \emph {et~al.}(2001)\citenamefont
  {{Ossendrijver}}, \citenamefont {{Stix}},\ and\ \citenamefont
  {{Brandenburg}}}]{Ossendrijver+2001}%
  \BibitemOpen
  \bibfield  {author} {\bibinfo {author} {\bibfnamefont {M.}~\bibnamefont
  {{Ossendrijver}}}, \bibinfo {author} {\bibfnamefont {M.}~\bibnamefont
  {{Stix}}},\ and\ \bibinfo {author} {\bibfnamefont {A.}~\bibnamefont
  {{Brandenburg}}},\ }\bibfield  {title} {\bibinfo {title} {{Magnetoconvection
  and dynamo coefficients:. Dependence of the alpha effect on rotation and
  magnetic field}},\ }\href {https://doi.org/10.1051/0004-6361:20011041}
  {\bibfield  {journal} {\bibinfo  {journal} {\aap}\ }\textbf {\bibinfo
  {volume} {376}},\ \bibinfo {pages} {713} (\bibinfo {year} {2001})},\ \Eprint
  {https://arxiv.org/abs/astro-ph/0108274} {arXiv:astro-ph/0108274 [astro-ph]}
  \BibitemShut {NoStop}%
\bibitem [{\citenamefont {{Tobias}}\ and\ \citenamefont
  {{Cattaneo}}(2013)}]{TobiasCattaneo2013}%
  \BibitemOpen
  \bibfield  {author} {\bibinfo {author} {\bibfnamefont {S.~M.}\ \bibnamefont
  {{Tobias}}}\ and\ \bibinfo {author} {\bibfnamefont {F.}~\bibnamefont
  {{Cattaneo}}},\ }\bibfield  {title} {\bibinfo {title} {{Shear-driven dynamo
  waves at high magnetic Reynolds number}},\ }\href
  {https://doi.org/10.1038/nature12177} {\bibfield  {journal} {\bibinfo
  {journal} {\nat}\ }\textbf {\bibinfo {volume} {497}},\ \bibinfo {pages} {463}
  (\bibinfo {year} {2013})}\BibitemShut {NoStop}%
\bibitem [{\citenamefont {{Pouquet}}\ \emph {et~al.}(1976)\citenamefont
  {{Pouquet}}, \citenamefont {{Frisch}},\ and\ \citenamefont
  {{Leorat}}}]{Pouquet+1976}%
  \BibitemOpen
  \bibfield  {author} {\bibinfo {author} {\bibfnamefont {A.}~\bibnamefont
  {{Pouquet}}}, \bibinfo {author} {\bibfnamefont {U.}~\bibnamefont
  {{Frisch}}},\ and\ \bibinfo {author} {\bibfnamefont {J.}~\bibnamefont
  {{Leorat}}},\ }\bibfield  {title} {\bibinfo {title} {{Strong MHD helical
  turbulence and the nonlinear dynamo effect}},\ }\href
  {https://doi.org/10.1017/S0022112076002140} {\bibfield  {journal} {\bibinfo
  {journal} {Journal of Fluid Mechanics}\ }\textbf {\bibinfo {volume} {77}},\
  \bibinfo {pages} {321} (\bibinfo {year} {1976})}\BibitemShut {NoStop}%
\bibitem [{\citenamefont {Kleeorin}\ and\ \citenamefont
  {Ruzmaikin}(1982)}]{Kleeorin+1982}%
  \BibitemOpen
  \bibfield  {author} {\bibinfo {author} {\bibfnamefont {N.}~\bibnamefont
  {Kleeorin}}\ and\ \bibinfo {author} {\bibfnamefont {A.}~\bibnamefont
  {Ruzmaikin}},\ }\bibfield  {title} {\bibinfo {title} {Dynamics of the average
  turbulent helicity in a magnetic field},\ }\href@noop {} {\bibfield
  {journal} {\bibinfo  {journal} {Magnetohydrodynamics}\ }\textbf {\bibinfo
  {volume} {18}},\ \bibinfo {pages} {116} (\bibinfo {year} {1982})}\BibitemShut
  {NoStop}%
\bibitem [{\citenamefont {{Blackman}}\ and\ \citenamefont
  {{Field}}(2000)}]{BlackmanField2000}%
  \BibitemOpen
  \bibfield  {author} {\bibinfo {author} {\bibfnamefont {E.~G.}\ \bibnamefont
  {{Blackman}}}\ and\ \bibinfo {author} {\bibfnamefont {G.~B.}\ \bibnamefont
  {{Field}}},\ }\bibfield  {title} {\bibinfo {title} {{Constraints on the
  Magnitude of {\ensuremath{\alpha}} in Dynamo Theory}},\ }\href
  {https://doi.org/10.1086/308767} {\bibfield  {journal} {\bibinfo  {journal}
  {\apj}\ }\textbf {\bibinfo {volume} {534}},\ \bibinfo {pages} {984} (\bibinfo
  {year} {2000})},\ \Eprint {https://arxiv.org/abs/astro-ph/9903384}
  {arXiv:astro-ph/9903384 [astro-ph]} \BibitemShut {NoStop}%
\bibitem [{\citenamefont {{Blackman}}\ and\ \citenamefont
  {{Field}}(2002)}]{BlackmanField2002}%
  \BibitemOpen
  \bibfield  {author} {\bibinfo {author} {\bibfnamefont {E.~G.}\ \bibnamefont
  {{Blackman}}}\ and\ \bibinfo {author} {\bibfnamefont {G.~B.}\ \bibnamefont
  {{Field}}},\ }\bibfield  {title} {\bibinfo {title} {{New Dynamical Mean-Field
  Dynamo Theory and Closure Approach}},\ }\href
  {https://doi.org/10.1103/PhysRevLett.89.265007} {\bibfield  {journal}
  {\bibinfo  {journal} {\prl}\ }\textbf {\bibinfo {volume} {89}},\ \bibinfo
  {eid} {265007} (\bibinfo {year} {2002})},\ \Eprint
  {https://arxiv.org/abs/astro-ph/0207435} {arXiv:astro-ph/0207435 [astro-ph]}
  \BibitemShut {NoStop}%
\bibitem [{\citenamefont {{Blackman}}(2003)}]{Blackman2003}%
  \BibitemOpen
  \bibfield  {author} {\bibinfo {author} {\bibfnamefont {E.~G.}\ \bibnamefont
  {{Blackman}}},\ }\bibfield  {title} {\bibinfo {title} {{Understanding helical
  magnetic dynamo spectra with a non-linear four-scale theory}},\ }\href
  {https://doi.org/10.1046/j.1365-8711.2003.06812.x} {\bibfield  {journal}
  {\bibinfo  {journal} {\mnras}\ }\textbf {\bibinfo {volume} {344}},\ \bibinfo
  {pages} {707} (\bibinfo {year} {2003})},\ \Eprint
  {https://arxiv.org/abs/astro-ph/0301432} {arXiv:astro-ph/0301432 [astro-ph]}
  \BibitemShut {NoStop}%
\bibitem [{\citenamefont {{Brandenburg}}\ \emph {et~al.}(2008)\citenamefont
  {{Brandenburg}}, \citenamefont {{R{\"a}dler}}, \citenamefont {{Rheinhardt}},\
  and\ \citenamefont {{Subramanian}}}]{Brandenburg+2008}%
  \BibitemOpen
  \bibfield  {author} {\bibinfo {author} {\bibfnamefont {A.}~\bibnamefont
  {{Brandenburg}}}, \bibinfo {author} {\bibfnamefont {K.-H.}\ \bibnamefont
  {{R{\"a}dler}}}, \bibinfo {author} {\bibfnamefont {M.}~\bibnamefont
  {{Rheinhardt}}},\ and\ \bibinfo {author} {\bibfnamefont {K.}~\bibnamefont
  {{Subramanian}}},\ }\bibfield  {title} {\bibinfo {title} {{Magnetic Quenching
  of {\ensuremath{\alpha}} and Diffusivity Tensors in Helical Turbulence}},\
  }\href {https://doi.org/10.1086/593146} {\bibfield  {journal} {\bibinfo
  {journal} {\apjl}\ }\textbf {\bibinfo {volume} {687}},\ \bibinfo {pages}
  {L49} (\bibinfo {year} {2008})},\ \Eprint {https://arxiv.org/abs/0805.1287}
  {arXiv:0805.1287 [astro-ph]} \BibitemShut {NoStop}%
\bibitem [{\citenamefont {{Brandenburg}}\ and\ \citenamefont
  {{Sarson}}(2002)}]{BrandenburgSarson2002}%
  \BibitemOpen
  \bibfield  {author} {\bibinfo {author} {\bibfnamefont {A.}~\bibnamefont
  {{Brandenburg}}}\ and\ \bibinfo {author} {\bibfnamefont {G.~R.}\ \bibnamefont
  {{Sarson}}},\ }\bibfield  {title} {\bibinfo {title} {{Effect of
  Hyperdiffusivity on Turbulent Dynamos with Helicity}},\ }\href
  {https://doi.org/10.1103/PhysRevLett.88.055003} {\bibfield  {journal}
  {\bibinfo  {journal} {\prl}\ }\textbf {\bibinfo {volume} {88}},\ \bibinfo
  {eid} {055003} (\bibinfo {year} {2002})},\ \Eprint
  {https://arxiv.org/abs/astro-ph/0110171} {arXiv:astro-ph/0110171 [astro-ph]}
  \BibitemShut {NoStop}%
\bibitem [{\citenamefont {{Candelaresi}}\ and\ \citenamefont
  {{Brandenburg}}(2013)}]{CandelaresiBrandenburg2013}%
  \BibitemOpen
  \bibfield  {author} {\bibinfo {author} {\bibfnamefont {S.}~\bibnamefont
  {{Candelaresi}}}\ and\ \bibinfo {author} {\bibfnamefont {A.}~\bibnamefont
  {{Brandenburg}}},\ }\bibfield  {title} {\bibinfo {title} {{Kinetic helicity
  needed to drive large-scale dynamos}},\ }\href
  {https://doi.org/10.1103/PhysRevE.87.043104} {\bibfield  {journal} {\bibinfo
  {journal} {\pre}\ }\textbf {\bibinfo {volume} {87}},\ \bibinfo {eid} {043104}
  (\bibinfo {year} {2013})},\ \Eprint {https://arxiv.org/abs/1208.4529}
  {arXiv:1208.4529 [astro-ph.SR]} \BibitemShut {NoStop}%
\bibitem [{\citenamefont {{Vishniac}}\ and\ \citenamefont
  {{Cho}}(2001)}]{VishniacCho2001}%
  \BibitemOpen
  \bibfield  {author} {\bibinfo {author} {\bibfnamefont {E.~T.}\ \bibnamefont
  {{Vishniac}}}\ and\ \bibinfo {author} {\bibfnamefont {J.}~\bibnamefont
  {{Cho}}},\ }\bibfield  {title} {\bibinfo {title} {{Magnetic Helicity
  Conservation and Astrophysical Dynamos}},\ }\href
  {https://doi.org/10.1086/319817} {\bibfield  {journal} {\bibinfo  {journal}
  {\apj}\ }\textbf {\bibinfo {volume} {550}},\ \bibinfo {pages} {752} (\bibinfo
  {year} {2001})},\ \Eprint {https://arxiv.org/abs/astro-ph/0010373}
  {arXiv:astro-ph/0010373 [astro-ph]} \BibitemShut {NoStop}%
\bibitem [{\citenamefont {{Subramanian}}\ and\ \citenamefont
  {{Brandenburg}}(2004)}]{SubramanianBrandenburg2004}%
  \BibitemOpen
  \bibfield  {author} {\bibinfo {author} {\bibfnamefont {K.}~\bibnamefont
  {{Subramanian}}}\ and\ \bibinfo {author} {\bibfnamefont {A.}~\bibnamefont
  {{Brandenburg}}},\ }\bibfield  {title} {\bibinfo {title} {{Nonlinear Current
  Helicity Fluxes in Turbulent Dynamos and Alpha Quenching}},\ }\href
  {https://doi.org/10.1103/PhysRevLett.93.205001} {\bibfield  {journal}
  {\bibinfo  {journal} {\prl}\ }\textbf {\bibinfo {volume} {93}},\ \bibinfo
  {eid} {205001} (\bibinfo {year} {2004})},\ \Eprint
  {https://arxiv.org/abs/astro-ph/0408020} {arXiv:astro-ph/0408020 [astro-ph]}
  \BibitemShut {NoStop}%
\bibitem [{\citenamefont {{Hubbard}}\ and\ \citenamefont
  {{Brandenburg}}(2012)}]{HubbardBrandenburg2012}%
  \BibitemOpen
  \bibfield  {author} {\bibinfo {author} {\bibfnamefont {A.}~\bibnamefont
  {{Hubbard}}}\ and\ \bibinfo {author} {\bibfnamefont {A.}~\bibnamefont
  {{Brandenburg}}},\ }\bibfield  {title} {\bibinfo {title} {{Catastrophic
  Quenching in {\ensuremath{\alpha}}{\ensuremath{\Omega}} Dynamos Revisited}},\
  }\href {https://doi.org/10.1088/0004-637X/748/1/51} {\bibfield  {journal}
  {\bibinfo  {journal} {\apj}\ }\textbf {\bibinfo {volume} {748}},\ \bibinfo
  {eid} {51} (\bibinfo {year} {2012})},\ \Eprint
  {https://arxiv.org/abs/1107.0238} {arXiv:1107.0238 [astro-ph.SR]}
  \BibitemShut {NoStop}%
\bibitem [{\citenamefont {{Rincon}}(2021)}]{Rincon2021}%
  \BibitemOpen
  \bibfield  {author} {\bibinfo {author} {\bibfnamefont {F.}~\bibnamefont
  {{Rincon}}},\ }\bibfield  {title} {\bibinfo {title} {{Helical turbulent
  nonlinear dynamo at large magnetic Reynolds numbers}},\ }\href
  {https://doi.org/10.1103/PhysRevFluids.6.L121701} {\bibfield  {journal}
  {\bibinfo  {journal} {Physical Review Fluids}\ }\textbf {\bibinfo {volume}
  {6}},\ \bibinfo {eid} {L121701} (\bibinfo {year} {2021})},\ \Eprint
  {https://arxiv.org/abs/2108.12037} {arXiv:2108.12037 [physics.flu-dyn]}
  \BibitemShut {NoStop}%
\bibitem [{\citenamefont {{Gopalakrishnan}}\ and\ \citenamefont
  {{Subramanian}}(2022)}]{GopalakrishnanSubramanian2022}%
  \BibitemOpen
  \bibfield  {author} {\bibinfo {author} {\bibfnamefont {K.}~\bibnamefont
  {{Gopalakrishnan}}}\ and\ \bibinfo {author} {\bibfnamefont {K.}~\bibnamefont
  {{Subramanian}}},\ }\bibfield  {title} {\bibinfo {title} {{Magnetic helicity
  fluxes from triple correlators}},\ }\href@noop {} {\bibfield  {journal}
  {\bibinfo  {journal} {arXiv e-prints}\ ,\ \bibinfo {eid} {arXiv:2209.14810}}
  (\bibinfo {year} {2022})},\ \Eprint {https://arxiv.org/abs/2209.14810}
  {arXiv:2209.14810 [astro-ph.GA]} \BibitemShut {NoStop}%
\bibitem [{\citenamefont {{Bhat}}(2022)}]{Bhat2022}%
  \BibitemOpen
  \bibfield  {author} {\bibinfo {author} {\bibfnamefont {P.}~\bibnamefont
  {{Bhat}}},\ }\bibfield  {title} {\bibinfo {title} {{Saturation of large-scale
  dynamo in anisotropically forced turbulence}},\ }\href
  {https://doi.org/10.1093/mnras/stab3138} {\bibfield  {journal} {\bibinfo
  {journal} {\mnras}\ }\textbf {\bibinfo {volume} {509}},\ \bibinfo {pages}
  {2249} (\bibinfo {year} {2022})},\ \Eprint {https://arxiv.org/abs/2108.08740}
  {arXiv:2108.08740 [astro-ph.SR]} \BibitemShut {NoStop}%
\bibitem [{\citenamefont {{Brandenburg}}\ and\ \citenamefont
  {{Subramanian}}(2005)}]{BrandenburgSubramanian2005}%
  \BibitemOpen
  \bibfield  {author} {\bibinfo {author} {\bibfnamefont {A.}~\bibnamefont
  {{Brandenburg}}}\ and\ \bibinfo {author} {\bibfnamefont {K.}~\bibnamefont
  {{Subramanian}}},\ }\bibfield  {title} {\bibinfo {title} {{Astrophysical
  magnetic fields and nonlinear dynamo theory}},\ }\href
  {https://doi.org/10.1016/j.physrep.2005.06.005} {\bibfield  {journal}
  {\bibinfo  {journal} {\physrep}\ }\textbf {\bibinfo {volume} {417}},\
  \bibinfo {pages} {1} (\bibinfo {year} {2005})},\ \Eprint
  {https://arxiv.org/abs/astro-ph/0405052} {arXiv:astro-ph/0405052 [astro-ph]}
  \BibitemShut {NoStop}%
\bibitem [{\citenamefont {{Brandenburg}}(2018)}]{Brandenburg2018}%
  \BibitemOpen
  \bibfield  {author} {\bibinfo {author} {\bibfnamefont {A.}~\bibnamefont
  {{Brandenburg}}},\ }\bibfield  {title} {\bibinfo {title} {{Advances in
  mean-field dynamo theory and applications to astrophysical turbulence}},\
  }\href {https://doi.org/10.1017/S0022377818000806} {\bibfield  {journal}
  {\bibinfo  {journal} {Journal of Plasma Physics}\ }\textbf {\bibinfo {volume}
  {84}},\ \bibinfo {eid} {735840404} (\bibinfo {year} {2018})},\ \Eprint
  {https://arxiv.org/abs/1801.05384} {arXiv:1801.05384 [physics.flu-dyn]}
  \BibitemShut {NoStop}%
\bibitem [{\citenamefont {{Hughes}}(2018)}]{Hughes2018}%
  \BibitemOpen
  \bibfield  {author} {\bibinfo {author} {\bibfnamefont {D.~W.}\ \bibnamefont
  {{Hughes}}},\ }\bibfield  {title} {\bibinfo {title} {{Mean field
  electrodynamics: triumphs and tribulations}},\ }\href
  {https://doi.org/10.1017/S0022377818000855} {\bibfield  {journal} {\bibinfo
  {journal} {Journal of Plasma Physics}\ }\textbf {\bibinfo {volume} {84}},\
  \bibinfo {eid} {735840407} (\bibinfo {year} {2018})},\ \Eprint
  {https://arxiv.org/abs/1804.02877} {arXiv:1804.02877 [astro-ph.SR]}
  \BibitemShut {NoStop}%
\bibitem [{\citenamefont {{Rincon}}(2019)}]{Rincon2019}%
  \BibitemOpen
  \bibfield  {author} {\bibinfo {author} {\bibfnamefont {F.}~\bibnamefont
  {{Rincon}}},\ }\bibfield  {title} {\bibinfo {title} {{Dynamo theories}},\
  }\href {https://doi.org/10.1017/S0022377819000539} {\bibfield  {journal}
  {\bibinfo  {journal} {Journal of Plasma Physics}\ }\textbf {\bibinfo {volume}
  {85}},\ \bibinfo {eid} {205850401} (\bibinfo {year} {2019})},\ \Eprint
  {https://arxiv.org/abs/1903.07829} {arXiv:1903.07829 [physics.plasm-ph]}
  \BibitemShut {NoStop}%
\bibitem [{\citenamefont {{Brandenburg}}\ and\ \citenamefont
  {{Ntormousi}}(2022)}]{BrandenburgNtormousi2022}%
  \BibitemOpen
  \bibfield  {author} {\bibinfo {author} {\bibfnamefont {A.}~\bibnamefont
  {{Brandenburg}}}\ and\ \bibinfo {author} {\bibfnamefont {E.}~\bibnamefont
  {{Ntormousi}}},\ }\bibfield  {title} {\bibinfo {title} {{Galactic Dynamos}},\
  }\href@noop {} {\bibfield  {journal} {\bibinfo  {journal} {arXiv e-prints}\
  ,\ \bibinfo {eid} {arXiv:2211.03476}} (\bibinfo {year} {2022})},\ \Eprint
  {https://arxiv.org/abs/2211.03476} {arXiv:2211.03476 [astro-ph.GA]}
  \BibitemShut {NoStop}%
\bibitem [{\citenamefont {{Subramanian}}(1999)}]{Subramanian1999}%
  \BibitemOpen
  \bibfield  {author} {\bibinfo {author} {\bibfnamefont {K.}~\bibnamefont
  {{Subramanian}}},\ }\bibfield  {title} {\bibinfo {title} {{Unified Treatment
  of Small- and Large-Scale Dynamos in Helical Turbulence}},\ }\href
  {https://doi.org/10.1103/PhysRevLett.83.2957} {\bibfield  {journal} {\bibinfo
   {journal} {\prl}\ }\textbf {\bibinfo {volume} {83}},\ \bibinfo {pages}
  {2957} (\bibinfo {year} {1999})},\ \Eprint
  {https://arxiv.org/abs/astro-ph/9908280} {arXiv:astro-ph/9908280 [astro-ph]}
  \BibitemShut {NoStop}%
\bibitem [{\citenamefont {{Boldyrev}}\ \emph {et~al.}(2005)\citenamefont
  {{Boldyrev}}, \citenamefont {{Cattaneo}},\ and\ \citenamefont
  {{Rosner}}}]{Boldyrev+2005}%
  \BibitemOpen
  \bibfield  {author} {\bibinfo {author} {\bibfnamefont {S.}~\bibnamefont
  {{Boldyrev}}}, \bibinfo {author} {\bibfnamefont {F.}~\bibnamefont
  {{Cattaneo}}},\ and\ \bibinfo {author} {\bibfnamefont {R.}~\bibnamefont
  {{Rosner}}},\ }\bibfield  {title} {\bibinfo {title} {{Magnetic-Field
  Generation in Helical Turbulence}},\ }\href
  {https://doi.org/10.1103/PhysRevLett.95.255001} {\bibfield  {journal}
  {\bibinfo  {journal} {\prl}\ }\textbf {\bibinfo {volume} {95}},\ \bibinfo
  {eid} {255001} (\bibinfo {year} {2005})},\ \Eprint
  {https://arxiv.org/abs/astro-ph/0504588} {arXiv:astro-ph/0504588 [astro-ph]}
  \BibitemShut {NoStop}%
\bibitem [{\citenamefont {{Bhat}}\ \emph {et~al.}(2016)\citenamefont {{Bhat}},
  \citenamefont {{Subramanian}},\ and\ \citenamefont
  {{Brandenburg}}}]{Bhat+2016}%
  \BibitemOpen
  \bibfield  {author} {\bibinfo {author} {\bibfnamefont {P.}~\bibnamefont
  {{Bhat}}}, \bibinfo {author} {\bibfnamefont {K.}~\bibnamefont
  {{Subramanian}}},\ and\ \bibinfo {author} {\bibfnamefont {A.}~\bibnamefont
  {{Brandenburg}}},\ }\bibfield  {title} {\bibinfo {title} {{A unified
  large/small-scale dynamo in helical turbulence}},\ }\href
  {https://doi.org/10.1093/mnras/stw1257} {\bibfield  {journal} {\bibinfo
  {journal} {\mnras}\ }\textbf {\bibinfo {volume} {461}},\ \bibinfo {pages}
  {240} (\bibinfo {year} {2016})},\ \Eprint {https://arxiv.org/abs/1508.02706}
  {arXiv:1508.02706 [astro-ph.GA]} \BibitemShut {NoStop}%
\bibitem [{\citenamefont {{Pietarila Graham}}\ \emph
  {et~al.}(2012)\citenamefont {{Pietarila Graham}}, \citenamefont {{Blackman}},
  \citenamefont {{Mininni}},\ and\ \citenamefont {{Pouquet}}}]{Graham+2012}%
  \BibitemOpen
  \bibfield  {author} {\bibinfo {author} {\bibfnamefont {J.}~\bibnamefont
  {{Pietarila Graham}}}, \bibinfo {author} {\bibfnamefont {E.~G.}\ \bibnamefont
  {{Blackman}}}, \bibinfo {author} {\bibfnamefont {P.~D.}\ \bibnamefont
  {{Mininni}}},\ and\ \bibinfo {author} {\bibfnamefont {A.}~\bibnamefont
  {{Pouquet}}},\ }\bibfield  {title} {\bibinfo {title} {{Not much helicity is
  needed to drive large-scale dynamos}},\ }\href
  {https://doi.org/10.1103/PhysRevE.85.066406} {\bibfield  {journal} {\bibinfo
  {journal} {\pre}\ }\textbf {\bibinfo {volume} {85}},\ \bibinfo {eid} {066406}
  (\bibinfo {year} {2012})},\ \Eprint {https://arxiv.org/abs/1108.3039}
  {arXiv:1108.3039 [physics.flu-dyn]} \BibitemShut {NoStop}%
\bibitem [{\citenamefont {{Bhat}}\ \emph {et~al.}(2019)\citenamefont {{Bhat}},
  \citenamefont {{Subramanian}},\ and\ \citenamefont
  {{Brandenburg}}}]{Bhat+2019}%
  \BibitemOpen
  \bibfield  {author} {\bibinfo {author} {\bibfnamefont {P.}~\bibnamefont
  {{Bhat}}}, \bibinfo {author} {\bibfnamefont {K.}~\bibnamefont
  {{Subramanian}}},\ and\ \bibinfo {author} {\bibfnamefont {A.}~\bibnamefont
  {{Brandenburg}}},\ }\bibfield  {title} {\bibinfo {title} {{Efficient
  quasi-kinematic large-scale dynamo as the small-scale dynamo saturates}},\
  }\href@noop {} {\bibfield  {journal} {\bibinfo  {journal} {arXiv e-prints}\
  ,\ \bibinfo {eid} {arXiv:1905.08278}} (\bibinfo {year} {2019})},\ \Eprint
  {https://arxiv.org/abs/1905.08278} {arXiv:1905.08278 [astro-ph.GA]}
  \BibitemShut {NoStop}%
\bibitem [{\citenamefont {{Brandenburg}}(2001)}]{Brandenburg2001}%
  \BibitemOpen
  \bibfield  {author} {\bibinfo {author} {\bibfnamefont {A.}~\bibnamefont
  {{Brandenburg}}},\ }\bibfield  {title} {\bibinfo {title} {{The Inverse
  Cascade and Nonlinear Alpha-Effect in Simulations of Isotropic Helical
  Hydromagnetic Turbulence}},\ }\href {https://doi.org/10.1086/319783}
  {\bibfield  {journal} {\bibinfo  {journal} {\apj}\ }\textbf {\bibinfo
  {volume} {550}},\ \bibinfo {pages} {824} (\bibinfo {year} {2001})},\ \Eprint
  {https://arxiv.org/abs/astro-ph/0006186} {arXiv:astro-ph/0006186 [astro-ph]}
  \BibitemShut {NoStop}%
\bibitem [{\citenamefont {{Bermudez}}\ and\ \citenamefont
  {{Alexakis}}(2022)}]{Bermudez+2022}%
  \BibitemOpen
  \bibfield  {author} {\bibinfo {author} {\bibfnamefont {G.}~\bibnamefont
  {{Bermudez}}}\ and\ \bibinfo {author} {\bibfnamefont {A.}~\bibnamefont
  {{Alexakis}}},\ }\bibfield  {title} {\bibinfo {title} {{Saturation of
  Turbulent Helical Dynamos}},\ }\href
  {https://doi.org/10.1103/PhysRevLett.129.195101} {\bibfield  {journal}
  {\bibinfo  {journal} {\prl}\ }\textbf {\bibinfo {volume} {129}},\ \bibinfo
  {eid} {195101} (\bibinfo {year} {2022})},\ \Eprint
  {https://arxiv.org/abs/2204.14091} {arXiv:2204.14091 [physics.flu-dyn]}
  \BibitemShut {NoStop}%
\bibitem [{\citenamefont {{Pencil Code Collaboration}}\ \emph
  {et~al.}(2021)\citenamefont {{Pencil Code Collaboration}}, \citenamefont
  {{Brandenburg}}, \citenamefont {{Johansen}}, \citenamefont {{Bourdin}},
  \citenamefont {{Dobler}}, \citenamefont {{Lyra}}, \citenamefont
  {{Rheinhardt}}, \citenamefont {{Bingert}}, \citenamefont {{Haugen}},
  \citenamefont {{Mee}}, \citenamefont {{Gent}}, \citenamefont {{Babkovskaia}},
  \citenamefont {{Yang}}, \citenamefont {{Heinemann}}, \citenamefont
  {{Dintrans}}, \citenamefont {{Mitra}}, \citenamefont {{Candelaresi}},
  \citenamefont {{Warnecke}}, \citenamefont {{K{\"a}pyl{\"a}}}, \citenamefont
  {{Schreiber}}, \citenamefont {{Chatterjee}}, \citenamefont
  {{K{\"a}pyl{\"a}}}, \citenamefont {{Li}}, \citenamefont {{Kr{\"u}ger}},
  \citenamefont {{Aarnes}}, \citenamefont {{Sarson}}, \citenamefont {{Oishi}},
  \citenamefont {{Schober}}, \citenamefont {{Plasson}}, \citenamefont
  {{Sandin}}, \citenamefont {{Karchniwy}}, \citenamefont {{Rodrigues}},
  \citenamefont {{Hubbard}}, \citenamefont {{Guerrero}}, \citenamefont
  {{Snodin}}, \citenamefont {{Losada}}, \citenamefont {{Pekkil{\"a}}},\ and\
  \citenamefont {{Qian}}}]{JOSS2021}%
  \BibitemOpen
  \bibfield  {author} {\bibinfo {author} {\bibnamefont {{Pencil Code
  Collaboration}}}, \bibinfo {author} {\bibfnamefont {A.}~\bibnamefont
  {{Brandenburg}}}, \bibinfo {author} {\bibfnamefont {A.}~\bibnamefont
  {{Johansen}}}, \bibinfo {author} {\bibfnamefont {P.}~\bibnamefont
  {{Bourdin}}}, \bibinfo {author} {\bibfnamefont {W.}~\bibnamefont {{Dobler}}},
  \bibinfo {author} {\bibfnamefont {W.}~\bibnamefont {{Lyra}}}, \bibinfo
  {author} {\bibfnamefont {M.}~\bibnamefont {{Rheinhardt}}}, \bibinfo {author}
  {\bibfnamefont {S.}~\bibnamefont {{Bingert}}}, \bibinfo {author}
  {\bibfnamefont {N.}~\bibnamefont {{Haugen}}}, \bibinfo {author}
  {\bibfnamefont {A.}~\bibnamefont {{Mee}}}, \bibinfo {author} {\bibfnamefont
  {F.}~\bibnamefont {{Gent}}}, \bibinfo {author} {\bibfnamefont
  {N.}~\bibnamefont {{Babkovskaia}}}, \bibinfo {author} {\bibfnamefont {C.-C.}\
  \bibnamefont {{Yang}}}, \bibinfo {author} {\bibfnamefont {T.}~\bibnamefont
  {{Heinemann}}}, \bibinfo {author} {\bibfnamefont {B.}~\bibnamefont
  {{Dintrans}}}, \bibinfo {author} {\bibfnamefont {D.}~\bibnamefont {{Mitra}}},
  \bibinfo {author} {\bibfnamefont {S.}~\bibnamefont {{Candelaresi}}}, \bibinfo
  {author} {\bibfnamefont {J.}~\bibnamefont {{Warnecke}}}, \bibinfo {author}
  {\bibfnamefont {P.}~\bibnamefont {{K{\"a}pyl{\"a}}}}, \bibinfo {author}
  {\bibfnamefont {A.}~\bibnamefont {{Schreiber}}}, \bibinfo {author}
  {\bibfnamefont {P.}~\bibnamefont {{Chatterjee}}}, \bibinfo {author}
  {\bibfnamefont {M.}~\bibnamefont {{K{\"a}pyl{\"a}}}}, \bibinfo {author}
  {\bibfnamefont {X.-Y.}\ \bibnamefont {{Li}}}, \bibinfo {author}
  {\bibfnamefont {J.}~\bibnamefont {{Kr{\"u}ger}}}, \bibinfo {author}
  {\bibfnamefont {J.}~\bibnamefont {{Aarnes}}}, \bibinfo {author}
  {\bibfnamefont {G.}~\bibnamefont {{Sarson}}}, \bibinfo {author}
  {\bibfnamefont {J.}~\bibnamefont {{Oishi}}}, \bibinfo {author} {\bibfnamefont
  {J.}~\bibnamefont {{Schober}}}, \bibinfo {author} {\bibfnamefont
  {R.}~\bibnamefont {{Plasson}}}, \bibinfo {author} {\bibfnamefont
  {C.}~\bibnamefont {{Sandin}}}, \bibinfo {author} {\bibfnamefont
  {E.}~\bibnamefont {{Karchniwy}}}, \bibinfo {author} {\bibfnamefont
  {L.}~\bibnamefont {{Rodrigues}}}, \bibinfo {author} {\bibfnamefont
  {A.}~\bibnamefont {{Hubbard}}}, \bibinfo {author} {\bibfnamefont
  {G.}~\bibnamefont {{Guerrero}}}, \bibinfo {author} {\bibfnamefont
  {A.}~\bibnamefont {{Snodin}}}, \bibinfo {author} {\bibfnamefont
  {I.}~\bibnamefont {{Losada}}}, \bibinfo {author} {\bibfnamefont
  {J.}~\bibnamefont {{Pekkil{\"a}}}},\ and\ \bibinfo {author} {\bibfnamefont
  {C.}~\bibnamefont {{Qian}}},\ }\bibfield  {title} {\bibinfo {title} {{The
  Pencil Code, a modular MPI code for partial differential equations and
  particles: multipurpose and multiuser-maintained}},\ }\href
  {https://doi.org/10.21105/joss.02807} {\bibfield  {journal} {\bibinfo
  {journal} {The Journal of Open Source Software}\ }\textbf {\bibinfo {volume}
  {6}},\ \bibinfo {eid} {2807} (\bibinfo {year} {2021})},\ \Eprint
  {https://arxiv.org/abs/2009.08231} {arXiv:2009.08231 [astro-ph.IM]}
  \BibitemShut {NoStop}%
\bibitem [{\citenamefont {{Zhou}}\ and\ \citenamefont
  {{Blackman}}(2023{\natexlab{a}})}]{zenodo}%
  \BibitemOpen
  \bibfield  {author} {\bibinfo {author} {\bibfnamefont {H.}~\bibnamefont
  {{Zhou}}}\ and\ \bibinfo {author} {\bibfnamefont {E.}~\bibnamefont
  {{Blackman}}},\ }\bibfield  {title} {\bibinfo {title} {{Dataset for ``Helical
  dynamo growth at modest versus extreme magnetic Reynolds numbers''}},\ }\href
  {https://doi.org/10.5281/zenodo.7632994} {10.5281/zenodo.7632994} (\bibinfo
  {year} {2023}{\natexlab{a}}),\ \bibinfo {note} {{Zenodo}}\BibitemShut
  {NoStop}%
\bibitem [{\citenamefont {{Zhou}}\ and\ \citenamefont
  {{Blackman}}(2023{\natexlab{b}})}]{sm}%
  \BibitemOpen
  \bibfield  {author} {\bibinfo {author} {\bibfnamefont {H.}~\bibnamefont
  {{Zhou}}}\ and\ \bibinfo {author} {\bibfnamefont {E.}~\bibnamefont
  {{Blackman}}},\ }\bibfield  {title} {\bibinfo {title} {{Supplemental Material
  for the manuscript ``Helical dynamo growth at modest versus extreme magnetic
  Reynolds numbers''}},\ }\href@noop {} {} (\bibinfo {year}
  {2023}{\natexlab{b}})\BibitemShut {NoStop}%
\bibitem [{\citenamefont {{Galishnikova}}\ \emph {et~al.}(2022)\citenamefont
  {{Galishnikova}}, \citenamefont {{Kunz}},\ and\ \citenamefont
  {{Schekochihin}}}]{Galishnikova+2022}%
  \BibitemOpen
  \bibfield  {author} {\bibinfo {author} {\bibfnamefont {A.~K.}\ \bibnamefont
  {{Galishnikova}}}, \bibinfo {author} {\bibfnamefont {M.~W.}\ \bibnamefont
  {{Kunz}}},\ and\ \bibinfo {author} {\bibfnamefont {A.~A.}\ \bibnamefont
  {{Schekochihin}}},\ }\bibfield  {title} {\bibinfo {title} {{Tearing
  Instability and Current-Sheet Disruption in the Turbulent Dynamo}},\ }\href
  {https://doi.org/10.1103/PhysRevX.12.041027} {\bibfield  {journal} {\bibinfo
  {journal} {Physical Review X}\ }\textbf {\bibinfo {volume} {12}},\ \bibinfo
  {eid} {041027} (\bibinfo {year} {2022})},\ \Eprint
  {https://arxiv.org/abs/2201.07757} {arXiv:2201.07757 [astro-ph.HE]}
  \BibitemShut {NoStop}%
\bibitem [{\citenamefont {{Beck}}\ \emph {et~al.}(2019)\citenamefont {{Beck}},
  \citenamefont {{Chamandy}}, \citenamefont {{Elson}},\ and\ \citenamefont
  {{Blackman}}}]{Beck+2019}%
  \BibitemOpen
  \bibfield  {author} {\bibinfo {author} {\bibfnamefont {R.}~\bibnamefont
  {{Beck}}}, \bibinfo {author} {\bibfnamefont {L.}~\bibnamefont {{Chamandy}}},
  \bibinfo {author} {\bibfnamefont {E.}~\bibnamefont {{Elson}}},\ and\ \bibinfo
  {author} {\bibfnamefont {E.~G.}\ \bibnamefont {{Blackman}}},\ }\bibfield
  {title} {\bibinfo {title} {{Synthesizing Observations and Theory to
  Understand Galactic Magnetic Fields: Progress and Challenges}},\ }\href
  {https://doi.org/10.3390/galaxies8010004} {\bibfield  {journal} {\bibinfo
  {journal} {Galaxies}\ }\textbf {\bibinfo {volume} {8}},\ \bibinfo {pages} {4}
  (\bibinfo {year} {2019})},\ \Eprint {https://arxiv.org/abs/1912.08962}
  {arXiv:1912.08962 [astro-ph.GA]} \BibitemShut {NoStop}%
\bibitem [{\citenamefont {{Skoutnev}}\ \emph {et~al.}(2022)\citenamefont
  {{Skoutnev}}, \citenamefont {{Squire}},\ and\ \citenamefont
  {{Bhattacharjee}}}]{Skoutnev+2022}%
  \BibitemOpen
  \bibfield  {author} {\bibinfo {author} {\bibfnamefont {V.}~\bibnamefont
  {{Skoutnev}}}, \bibinfo {author} {\bibfnamefont {J.}~\bibnamefont
  {{Squire}}},\ and\ \bibinfo {author} {\bibfnamefont {A.}~\bibnamefont
  {{Bhattacharjee}}},\ }\bibfield  {title} {\bibinfo {title} {{On large-scale
  dynamos with stable stratification and the application to stellar radiative
  zones}},\ }\href {https://doi.org/10.1093/mnras/stac2676} {\bibfield
  {journal} {\bibinfo  {journal} {\mnras}\ }\textbf {\bibinfo {volume} {517}},\
  \bibinfo {pages} {526} (\bibinfo {year} {2022})},\ \Eprint
  {https://arxiv.org/abs/2203.01943} {arXiv:2203.01943 [astro-ph.SR]}
  \BibitemShut {NoStop}%
\bibitem [{\citenamefont {{Biskamp}}\ \emph {et~al.}(1998)\citenamefont
  {{Biskamp}}, \citenamefont {{Schwarz}},\ and\ \citenamefont
  {{Celani}}}]{Biskamp+1998}%
  \BibitemOpen
  \bibfield  {author} {\bibinfo {author} {\bibfnamefont {D.}~\bibnamefont
  {{Biskamp}}}, \bibinfo {author} {\bibfnamefont {E.}~\bibnamefont
  {{Schwarz}}},\ and\ \bibinfo {author} {\bibfnamefont {A.}~\bibnamefont
  {{Celani}}},\ }\bibfield  {title} {\bibinfo {title} {{Nonlocal Bottleneck
  Effect in Two-Dimensional Turbulence}},\ }\href
  {https://doi.org/10.1103/PhysRevLett.81.4855} {\bibfield  {journal} {\bibinfo
   {journal} {\prl}\ }\textbf {\bibinfo {volume} {81}},\ \bibinfo {pages}
  {4855} (\bibinfo {year} {1998})},\ \Eprint
  {https://arxiv.org/abs/chao-dyn/9807012} {arXiv:chao-dyn/9807012 [nlin.CD]}
  \BibitemShut {NoStop}%
\end{thebibliography}%

\end{document}